\documentstyle[12pt]{article}
\def\appendix#1{
\addtocounter{section}{1}
\setcounter{equation}{0}
\renewcommand{\thesection}{\Alph{section}}
\section*{Appendix \thesection\protect\indent #1}
\addcontentsline{toc}{section}{Appendix \thesection\ \ \ #1}
}
\newcommand{\tr}[1]{\,{\rm tr}\,#1\,}

\def \N {{\cal N}}
\def \ov {\over}

\def\be{\begin{equation}}
\def\la{\label}
\def\ee{\end{equation}}
\def\bea{\begin{eqnarray}}
\def\eea{\end{eqnarray}}

\def\a{\alpha}
\def\b{\beta}
\def\s{\sigma}

\def\g{\gamma}
\def\d{\delta}

\def\k{\kappa}
\def\l{\left(}
\def\r{\right)}
\def\p{\partial}

\begin{document}
\title{\hfill{\small UAHEP995} \\
\vspace{-0.3cm}
\hfill{\small OHSTPY-HEP-T-99-023}\\
\vspace{0.5cm}
Three-point correlators of stress tensors\\ in 
maximally-supersymmetric
conformal theories\\ in $d=3$ and $d=6$}
\author{\large 
F. Bastianelli$^{a}$,
\mbox{} S. Frolov$^{b,}$\thanks{Also at Steklov Mathematical 
Institute, Moscow.}
\mbox{} and \mbox{} A.A. Tseytlin$^{c,}$\thanks{Also at Lebedev 
Physics Institute, Moscow and Imperial College, London.}
\mbox{}
\vspace{0.2cm} \\ \small  
$^a$Dipartimento di Fisica, Universit\`a di Bologna,
\vspace{-0.2cm} \mbox{} \\ \small  
V. Irnerio 46, I-40126 Bologna, Italy
\vspace{-0.25cm} \mbox{} \\ \small 
and
\vspace{-0.25cm} \mbox{} \\ \small 
INFN, Sezione di Bologna
\vspace{0.15cm}
\mbox{} \\ \small 
$^b$Department of Physics and Astronomy
\vspace{-0.2cm} \mbox{} \\ \small 
University of Alabama, Box 870324
\vspace{-0.2cm} \mbox{} \\ \small 
Tuscaloosa, Alabama 35487-0324, USA
\vspace{0.15cm}
\mbox{} \\ \small 
$^c$Department of Physics 
\vspace{-0.2cm} \mbox{} \\ \small 
The Ohio State University
\vspace{-0.2cm} \mbox{} \\ \small 
Columbus, OH 43210-1106
\mbox{}
}
\date {}
\maketitle
\vspace{-1cm}
\begin{abstract}
\vspace{-0.1cm}
We consider free superconformal theories of $\N=8$ scalar 
multiplet in $d=3$ and $(2,0)$ tensor multiplet 
in $d=6$ and compute 2-point and 3-point correlators of their 
stress  tensors.  The results  for the  2-point and  the  3-point 
correlators for a single $d=3$ and $d=6$  multiplet differ
from the ``strong-coupling" 
$AdS_4$ and $AdS_7$ supergravity predictions by 
the factors  ${4 \sqrt 2  \over 3 \pi} N^{3/2}$  and  $4N^3$ 
respectively. 
These are the same factors 
 as found  earlier in hep-th/9703040 in the comparison 
of the  brane  free field theory  and the $d=11$ 
supergravity predictions for the 
absorption cross-sections of longitudinally polarized 
gravitons 
by  $N$ \ M2 and M5 branes. 
While the correspondence of the  results for the
cross-sections and  
2-point functions was expected on the basis of unitarity, 
the fact that the same coefficients appear
in the ratio of the
free-theory and supergravity 
 3-point functions is non-trivial. 
Thus, like in the  $d=4$ SYM case, in both $d=3$ and $d=6$ 
theories the ratio
of the 3-point and 2-point correlators $<TTT>/<TT>$
is exactly the same in the free field theory 
and in the interacting CFT as described 
 (to leading order in
large $N$) by the 11-dimensional supergravity on
$AdS_{d+1} \times S^{10-d}$. 
\end{abstract}
\baselineskip 18pt
\section{Introduction and summary}
In contrast to the low-energy theory  on multiple D3-branes
represented by $\N=4$ SYM theory, 
the $d=3$ and $d=6$ superconformal theories describing a large number $N$ of 
coincident  M2 and M5 branes remain poorly understood
(see, e.g., \cite{aah,sei} for reviews and references).
The M2 brane CFT  is expected to be described  
by an IR fixed point 
of $d=3$ 
$U(N)$ SYM theory. The  coincident M5 brane theory is 
expected to be  a new kind of $d=6$ CFT 
 related 
  to  a theory of ``tensionless strings" \cite{strom} 
(for its  DLCQ description see   \cite{ahh}).
At a generic point of moduli space, i.e. away from 
the M5-brane  coincidence point, the M5 brane  theory  should be 
represented by  $N$ interacting tensor multiplets
(see  also  \cite{gan} and refs. therein).
Both $d=3$ and $d=6$ theories become free in the $N=1$ limit, but 
for $N > 1$ they are still lacking a computationally 
useful description. 
    
A few basic facts that  follow from the properties  
of the corresponding classical supergravity solutions 
\cite{duff,guv}
are: 

(i) The collective coordinates of a single M2 brane
are represented by 
$\N=8, d=3$ scalar multiplet
(8 scalars and 8 Majorana spinors) \cite{berg}, 
 while
the collective coordinates of a single M5 brane 
by  the (2,0) $d=6$ multiplet (5 scalars, 2 Weyl spinors and 
an anti-selfdual  2-form) \cite{gib}. 

(ii) While the entropy of $N$\   D3 branes scales 
in the usual  $N^2$ way \cite{peet},
the entropy of multiple  M5 branes scales  as $N^{3}$
and the entropy of M2 branes as $N^{3/2}$ \cite{kt}.
This suggests that there is a corresponding enhancement of the
number  of light degrees of freedom when branes are put together.
The guidance of the supergravity solution   should probably 
be trusted more in the M5-brane  case  which is a 
 non-singular background. This is the case we shall mostly 
concentrate on in what follows. 
 The  M-theory  $R^4$
correction to  the $d=11$  supergravity action
results \cite{gktt} in a subleading $O(N)$ correction to the M5-brane 
entropy.  
 
(iii) The same $N^{3/2}$ and $N^3$ scalings dictated by 
the $d=11$ supergravity 
description are  found \cite{kl,GKT}
in the absorption rate of longitudinally polarized 
gravitons by M2-branes and M5-branes. 
The  supergravity absorption cross-sections 
 have the same form \cite{kl} 
as the cross-sections  in the  free theories of 
$d=3$ and  $d=6$  multiplets,    but, in addition to 
the $N^{3/2}$ and
$N^3$  factors, the supergravity
 and the free field theory predictions  
 differ also in the 
numerical coefficients \cite{GKT}\footnote{The hep-th version 
of \cite{GKT} used the  M2 absorption cross-section which 
was off by the factor of
8  compared to  the result of  \cite{emp}. This 
was corrected in the published  NPB version of \cite{GKT}.}
\be
{\sigma_{2\ sugra} \over \sigma_{2\ free \ f.t.} } 
= { 4 \sqrt 2 \ov 3 \pi } N^{3/2} 
\ , \ \ \ \ \ \ \ \ 
{\sigma_{5\ sugra} \over \sigma_{5\ free \ f.t.}  }
= 4 N^{3} \ . 
\ee
This numerical discrepancy was absent in the D3 brane case 
where the precise agreement \cite{kl} between the 
free-theory and the supergravity absorption rates
can be understood \cite{gkl}  as a consequence of the 
non-renormalization
theorem  for the correlator of the two stress tensors
in $\N=4$ SYM theory,   and is
intimately  related \cite{gkp}
to the AdS/CFT correspondence \cite{mal,gkp,wit}.

Some 
$AdS_7\times S^4$ supergravity predictions for 
the   properties (spectrum, correlators, conformal anomaly) 
 of the (2,0) non-abelian tensor
multiplet theory were studied also in  \cite{ozz,min,lei,hall,
sken,harv,awata, with,corr,niew,BZ,fior,nish}.

Our prime  aim  below   is to compute the 2- and 3-point 
correlators  of the stress tensors in the free $d=6$  
tensor multiplet theory and to compare the result 
to the AdS supergravity correlators  found in \cite{LT} ($<TT>$) 
and \cite{AF} ($<TTT>$). Since the classical supergravity 
absorption cross-section is expected 
to be related to the AdS supergravity correlator \cite{gkp} and since 
the unitarity relates the  free field  theory 
graviton absorption amplitude
to the imaginary part of the (Minkowski-space) 
correlator of the two  stress tensors, 
one should  find the same  $4N^3$ result 
  for ratio of  the free-theory CFT and AdS  2-point  correlators.
We shall indeed confirm this  by  the  explicit computation.

Moreover, we shall find that the 3-point AdS  graviton correlators 
are again  reproduced  $exactly$
 by the $d=6$ free theory  $<TTT>$  correlator, 
 up to  the $same$ overall coefficient $4N^3$! Let us 
note   that the comparison of the 3-point 
correlators goes way beyond the absorption
calculations in \cite{GKT} -- the AdS/CFT approach allows us 
to compute the multiple graviton (stress tensor)
 correlators in a systematic
way, something  that   is  hard to do  in the context 
of  the standard 
classical absorption calculations.

Similar conclusion will be reached in the case of the $d=3$ theory:
 the ratios of the AdS and free $d=3$ CFT  predictions 
 for $<TT>$ and $<TTT>$   will be   again exactly 
 the  same as in (1.1) --  
$
{ 4\sqrt 2  \ov 3 \pi } N^{3/2} $.

In the absence of a free coupling parameter,  
and thus of a ``non-renormalization theorem'' argument 
 used  in  the $d=4$ ($\N=4$ SYM) case,  
one could expect that the 3-point correlators  may   have  
{\it different}  structures
 in the free-field  $d=6$ theory  and 
in  the `true' 
strongly coupled $d=6$ CFT   represented 
 (to the leading order in large $N$)   
 by the   $d=11$  supergravity on $AdS_7\times S^4$. 
That this does not happen seems to be 
  non-trivial. One plausible   explanation is 
that the requirements 
of maximal superconformal symmetry are   powerful  
enough to constrain the form of $<TTT>$ 
so that it is reproduced by the free theory calculation,  
up to the overall coefficient determined
 by the  coefficient in  $<TT>$
 (cf. \cite{par}).
  A possible point of view is 
 that this  coefficient 
 may  be $N$-dependent and should interpolate
 between the single free  multiplet field theory result 
 for $N=1$  and the non-trivial  CFT (supergravity) 
 result for $N \to \infty$.\footnote{One could try to 
  argue, as in  \cite{bangr}, that the structure of 
   the M-theory  action on $AdS_7 \times S^4$ implies 
    that both $<TT>$ and $<TTT>$ 
     should not receive subleading $1/N$ corrections:
     higher-order  
     corrections like  $R^4$ (written 
      in a specific ``Weyl-tensor" scheme
      \cite{gktt})  may  not change the expressions
     for the 2-graviton and 3-graviton $AdS$ correlators.}
 As far as the 2-point and 3-point correlators 
of  stress tensor multiplet states are concerned,
 the  mysterious  strongly coupled $d=6$ CFT
and the free $d=6$ tensor multiplet theory thus 
happen to be in the 
same ``universality class".

The common 
overall  numerical coefficient $4 N^3$ 
 of $<TT>$ and $<TTT>$
should
 have an important   meaning. 
 Since the only  $d=6$ CFT  we explicitly know
 is the free tensor multiplet, we may  try to ``model"
 the  CFT predicted, via  AdS/CFT correspondence,
   by the supergravity M5 brane description
 by starting with
  a number  of 
 free tensor multiplets and assigning internal indices to them. 
 In the $d=4$ case  the theory of free 
 $N^2$\  $\N=4$ vector multiplets indeed  reproduces the supergravity 
predictions
 for protected 2- and 3-point functions.
 The idea is then to try to fix the internal index structure of the 
 tensor multiplet theory using the AdS supergravity
 results as a guide.
  The supergravity $N^3$ scaling
  may be formally   reproduced   by a 
  ``model (2,0)  theory''  where the  2-tensor, scalar and spinor
   fields   carry
    $three$ 
   internal indices $i,j,k=1,...,N$,
  i.e.  the  (selfdual) 2-form  field   strength  is 
$H^{ijk}_{\mu\nu\lambda}$, etc.  
This theory may  be describing the case 
when $three$ (as opposed to two  \cite{strom}) 
M5-branes 
are simultaneously  put together.

 The entropy and correlators 
of  composite operators  like the stress tensor 
 would then scale as $N^3$ for 
large $N$ (assuming there is no symmetry in 
 internal indices which would 
reduce the coefficient $N^3$ by an 
integer factor).
  Since the field strength has non-zero  dimension, 
introducing interactions would break classical conformal invariance.
One may   speculate 
 that there may   exist a  
new   interacting   $d=6$  theory
based on $H^{ijk}_{\mu\nu\lambda}$ (plus its superpartners)
which is conformal at the $quantum$ level. 
The hope is then that a
 non-trivial quantum dynamics should  be responsible 
 for the remaining factor of  $4$   mismatch
 between the free theory and the AdS supergravity 
 predictions.\footnote{
The assignment  of the three  internal indices 
to the antisymmetric tensor seems  suggested by   the  
heuristic explanation of the $N^3$ growth 
of the M5-brane entropy  as being due to  triple 
M5-brane connections by membranes of `pants' shape
(and is also related 
 to the presence of the cubic
  $\int C_3 dC_3 dC_3 $ term in the 11-d 
supergravity action, cf. \cite{harv}).
Virtual triple  connections  are 
  not dominant  in the case of  the open strings
  ending on  D-branes 
 (the 3-string interactions are subleading in the coupling)
but are very natural for M5-branes connected by membranes:
any membrane surface ending 
on several M5-branes may be cut into  `pants', with 
pair-wise (cylinder) connections being subleading  at large $N$ 
compared to the triple ones.  
The importance of similar triple M5 brane connections by 
 membranes   with 3 
boundaries was suggested in \cite{kttt} 
  in order to  explain the scaling 
  of the  entropy of the extremal 4-d black hole
 described by the $2555$ intersecting 
  M-brane  configuration
($ S \sim  \sqrt {N_1N_2N_3}$ where $N_i$ are charges of M5 branes).} 
Ignoring selfduality (and supersymmetry) constraints, 
 it seems likely    that  if  a consistent 
  interacting   theory of non-abelian  antisymmetric
tensors in 6 dimensions $B^{ijk}_{\mu\nu}$
exists, its action should contain infinite number of terms
with leading interaction  being quartic in $B^{ijk}_{\mu\nu}$,  
and  with $H^{ijk}_{\mu\nu\lambda}$ playing the role of an effective 
``coupling constant'' or the  structure tensor  of the corresponding 
``soft'' gauge algebra.\footnote{This may  probably allow 
to avoid 
the ``no-go'' theorem  of  \cite{hano}.}
Conformal invariance  at the quantum level  may be possible to achieve 
provided the dimension of $B^{ijk}_{\mu\nu}$
 is shifted from its classical value 2.\footnote{Such 
interacting  non-abelian  
(2,0) supersymmetric tensor 
multiplet theory  may be 
  a low-energy limit of a
    kind of ``tensionless $d=6$  string theory''. 
At a very   speculative level, 
one may think of  closed strings in $d=6$ with $three$
``Chan-Paton'' indices
which may originate from virtual membranes connecting $three$
parallel M5-branes. When the distances between 
M5-branes reduce to zero, the membranes with 3 holes may 
 produce a  special kind of strings
 which somehow carry three  internal indices
   (they may be visualized as 
    `blown-up' 3-string junctions,  or ``triangles").
 The basic interaction at the boundary  may  then be  described by 
$B^{ijk}_{\mu\nu}$.}
 
The structure of the paper is  the  following. 
To compute the free-theory correlators
$<TT>$ and $<TTT>$ in (2,0) theory  one needs to 
sum together the independent 2-form, scalar and spinor stress tensor 
contributions.  The 2- and 3-point functions of stress tensors
of the conformal scalar and spinor fields 
in an arbitrary dimension $d$ were already computed in 
\cite{OP} (and refs. therein).
The free $k$-form field theory is conformal in dimension 
$d=2k+2$.  In section 2 we present the {general} 
results  for the correlators of  
 two and three stress tensors in such  
 theory. Particular cases include 
the previously known $d=4$ vector case and the new $d=6$ 2-form 
case we are  interested in. Our general $d=2k+2$ 
results may be useful  in other contexts.

In section 3 we specialize to the case of the (2,0) tensor 
multiplet in $d=6$. To find the contribution
of the chiral 2-form field  we employ the method similar to the one
used in \cite{AW,GKT} which is based on the use 
of the non-chiral 2-form propagator  with the chirality projectors 
being inserted in the ``vertices" (composite operators).
Feynman rules obtained from actions for chiral bosons
\cite{HenTei,PST,Van}
should  produce equivalent results, as it happens
in the case of gravitational anomalies \cite{PvN,Lech}.
We find that the   3-point correlator of the
 total free-theory
stress 
tensor  has  exactly the same form as found in \cite{AF}
from the $AdS_7$
supergravity description, apart from the overall coefficient.
The latter coefficient  $4N^3$ 
 is the same as in the  2-point function.
The overall scale of $<TTT>$ is, in fact, 
 related to that of $<TT>$ by  the conformal Ward identity.

In section 4 we repeat the same computation
 in the case of the free 
theory of 8 scalars and 8 Majorana spinors  in $d=3$ 
and show that again the free field theory  3-point function 
is the same as the $AdS_4$ one, apart from the overall
coefficient which is the same
 as  in the ratio of the 2-point functions 
or as  in (1.1). 

Thus, like in   the $d=4$ SYM case, 
in both $d=3$ and $d=6$ 
cases the ratio
of the 3-point and 2-point stress tensor correlators 
is exactly the same in the free superconformal
field theory 
and in  the interacting CFT as described 
 to leading order in
large $N$ by the 11-dimensional supergravity on
$AdS_{d+1} \times S^{10-d}$, 
\be 
 \l {<TTT>\over <TT> } \r_{free\ f.t.} 
 = \l {<TTT>\over <TT> } \r_{sugra} \ . 
\ee
We expect  that, as  in the D3-brane case \cite{minw}, 
similar results should 
hold also for all 2- and 3-point correlation functions
of states belonging to the short multiplet of the 
stress tensor, i.e. all such correlators 
 should be reproduced by the free field theory, 
apart from the  same overall normalization factors.

\section{ Free conformal theory of $k$-form field  in $d=2k+2$}
We begin by considering a free nonchiral $k$-form field 
theory.
The free field theory of a $k$-form $B_{\mu_1 \dots \mu_k}$
in the Minkowski space of $d=2k+2$ dimensions is described by 
the following action 
\be
S=   - {1\over 2 (k+1)!} \int d^d x\ 
H_{\mu_1 \dots \mu_{k+1}} H^{\mu_1 \dots \mu_{k+1}} 
\ , \la{act}
\ee
where $H_{\mu_1 \dots \mu_{k+1}} = \p_{\mu_1} B_{\mu_2 
\dots \mu_{k+1}}
 \pm   {\rm cyclic\ permutations}$.

It is well known that due to the gauge invariance the number of 
physical degrees of freedom in the model 
is $$n_{{\rm ph}}=\frac{(2k)!}{(k!)^2}\ .$$
After coupling this theory to gravity in the minimal way  
one can check its  Weyl invariance, which guarantees the 
conformal invariance 
in the flat space limit. Defining the stress-tensor as 
$T^{\mu\nu} = {2\over \sqrt{-g}} {\delta S\over \delta g_{\mu \nu}}$
one gets
\be
T_{\lambda \nu} = {1\over k!} 
H_{\lambda  \mu_1 \dots \mu_{k}} H_\nu{}^{\mu_1 \dots \mu_{k}} 
-{1\over 2(k+1)!} \eta_{\lambda \nu} H^2 , 
\la{T}
\ee
which is obviously traceless.

Consider  correlation functions of physical (gauge-invariant) 
observables which depend only on the field strength. To calculate them 
 we need to know the propagator of 
the field strength. The simplest way to find it
is to add to the action (\ref{act}) the Lorentz gauge-fixing term 
$\frac{1}{2(k-1)!}(\p^{\mu_1}   B_{\mu_1 \mu_2 \dots \mu_k})^2$. Then 
the propagator of the $k$-form  field  is given by
\bea
\langle B_{\mu_1 \dots \mu_k}(x) B^{\nu_1 \dots \nu_k}(y) \rangle =
{\alpha_d k!\over (x-y)^{d-2}} \delta_{\mu_1\dots\mu_k}^{\nu_1\dots\nu_k} \ , 
\nonumber
\eea
where
\bea
\alpha_d &=& {1\over { (d-2) \omega_{d-1}}} \ , \cr
\omega_{d-1} &=& {2 \pi^{d\over2} \over \Gamma({d\over2})}\ \ \ \ \ \ \ \ \ \
{\rm (volume\ of\ unit\ sphere\ } S^{d-1}) \ ,  \cr
 \delta_{\mu_1\dots\mu_k}^{\nu_1\dots\nu_k} &=& \delta_{[\mu_1}^{\nu_1}
\cdots \delta_{\mu_k]}^{\nu_k} \ , 
\nonumber
\eea
with $[\dots]$ denoting antisymmetrization with unit strength.
One then   gets the following  propagator 
for the gauge invariant field strength $H$
\be
\langle H_{\mu_1 \dots \mu_{k+1}}(x) H^{\nu_1 \dots \nu_{k+1}}(y) \rangle =
{\alpha_d k! (d-2) (k+1)^2 \over r^{d}} \biggl [
\delta_{\mu_1\dots\mu_{k+1}}^{\nu_1\dots\nu_{k+1}}
-\ d \ \hat r_{[\mu_1} \hat r^{[\nu_1}  
\delta_{\mu_2\dots\mu_{k+1}]}^{\nu_2\dots\nu_{k+1}]}\biggr ]
\la{prop}
\ee
with $$ r^\mu = x^\mu - y^\mu\ ,
 \ \ \ \ \ \ \ \ \
 \hat r^\mu = {r^\mu \over |r|}\ .  $$
As  is well known (see, e.g., \cite{OP} and refs. therein),
 the   two-point function of the stress tensor 
in a $d$-dimensional conformal 
field theory is fixed by the conformal invariance up to a constant, 
and can be represented in the form  
\bea
\langle T_{\a\b}(x)T_{\g\d}(y) 
\rangle=\frac{C_T}{r^{2d}}\ 
{\cal I}_{\a\b ,\g\d }(r)\ , \label{two}
\eea
where $$ {\cal I}_{\a\b ,\g\d } \equiv  \frac12 J_{\a\g}J_{\b\d}
+\frac12 J_{\a\d}J_{\b\g}-\frac 1d \d_{\a\b}\d_{\g\d}\  , 
\ \ \ \ \ \ \  
J_{\a\b}\equiv \d_{\a\b}-2\frac{r_\a r_\b}{r^2}\ . $$
Thus all we need to know is the constant $C_T$. To find it we consider the
correlators with the indices 1 and 2, and choose $y=0$ and $x_\a=\d_{1\a}$.
Then a straightforward computation gives\footnote{After this
paper was submitted
for publication we were informed that the 2-point function 
for the antisymmetric tensors was computed previously 
in
\cite{AN}, with  the equivalent result.}
\be
C_T = {1\over \omega_{d-1}^2}
{d^2\over 2} {(2k)!\over (k!)^2 } = {1\over \omega_{d-1}^2}
{d^2\over 2}n_{{\rm ph}} .
\la{ct}
\ee
As was shown in \cite{OP},  the 3-point 
function of the stress tensor 
in a conformal field theory is 
parametrized  
by three independent 
constants, and can be written in the form \cite{EO}
\bea 
\nonumber
T_{\a\b ,\g\d ,\rho\s}&=&\frac{1}{|x-y|^{d}|y-z|^{d}|x-z|^{d}}
\nonumber\\
\times &\biggl [&E_{\a\b, \a '\b '}E_{\g\d ,\g '\d '}E_{\rho\s ,\rho '\s '}
\biggl( {\cal A} J_{\a '\g '}(x-y)J_{\d '\rho '}(y-z)J_{\b '\s '}(z-x) 
\biggr.   \nonumber \\ \nonumber
&+& \biggl. {\cal B}J_{\a '\g '}(x-y)J_{\d '\rho '}(x-z)Y_{\b '}Z_{\s '}(y-z)^2
+cycl.\ perm. \biggr) \\
\nonumber
&+&{\cal C}\l {\cal I}_{\a\b ,\g\d } (x-y) \l 
\frac{Z_\rho Z_\s}{Z^2}-\frac{1}{d}\d_{\rho\s}\r
+cycl.\ perm.\r \\
\nonumber
&+&{\cal D}\l E_{\a\b ,\a '\b '}E_{\g\d , \g '\d '}X_{\a '}
Y_{\g '}(x-y)^2
J_{\b '\d '}(x-y)\l \frac{Z_\rho Z_\s}{Z^2}-\frac{1}{d}\d_{\rho\s}\r
+cycl.\ perm.\r  \\
\label{three}
&+&
{\cal E}
\l \frac{X_\a X_\b}{X^2}-\frac{1}{d}\d_{\a\b}\r
 \l \frac{Y_\g Y_\d}{Y^2}-\frac{1}{d}\d_{\g\d}\r
 \l \frac{Z_\rho Z_\s}{Z^2}-\frac{1}{d}\d_{\rho\s}\r \biggr ], 
\eea 
where
 $$E_{\a\b, \g\d}= \frac12 \d_{\a\g}\d_{\b\d}
+\frac12 \d_{\a\d}\d_{\b\g}-\frac 1d \d_{\a\b}\d_{\g\d}$$ is the traceless
symmetric projector and 
 $X_\a$, $Y_\a$, $Z_\a$ are the 
conformal vectors 
$$X_\a =\frac{(x-z)_\a}{(x-z)^2}-\frac{(x-y)_\a}{(x-y)^2},\quad
Y_\a =\frac{(y-x)_\a}{(y-x)^2}-\frac{(y-z)_\a}{(y-z)^2},\quad
Z_\a =\frac{(z-y)_\a}{(z-y)^2}-\frac{(z-x)_\a}{(z-x)^2}\ . $$
There are two linear relations \cite{OP,EO} between the 5  constants 
$\cal A,B,C,D,E$ entering (\ref{three}) 
that allow to express two of them 
in terms of  the remaining three
$$(d^2-4){\cal A} + (d+2){\cal B} -4d{\cal C}-2{\cal D}=0\ , $$
\be (d-2)(d+4){\cal B} -2d(d+2){\cal C}+8{\cal D}-4{\cal E}=0\ . \ee
We choose ${\cal A}$, ${\cal B}$ and ${\cal C}$ as the three 
independent constants. To find them we take 
$z=0,~y_\alpha =\delta_{\alpha 1},~x_\alpha =2\delta_{\alpha 1}$ and 
consider the following three correlators  
\bea
&&\langle T_{12}(x)T_{13}(y)T_{23}(z) \rangle =\ 2^{-d}\ \tau
\nonumber\\
&&\langle T_{23}(x)T_{24}(y)T_{34}(z) \rangle =\ 2^{-d}\ t
\nonumber\\
&&\langle T_{12}(x)T_{12}(y)T_{22}(z) \rangle =\ 2^{-d}\ (\rho +2\tau )
\ , \nonumber
\eea
where $\tau,~t,~\rho$ are the coefficients in the collinear frame from 
eq.(4.21) of \cite{OP}. They are related to ${\cal A}$, ${\cal B}$ and 
${\cal C}$ as follows (see eqs.(4.25) and (3.21) from \cite{OP}, 
and the footnote on p.21 of \cite{EO})
\bea
&&{\cal A}=8t
\la{A}\\
&&{\cal B}=8(\tau +t)
\la{B}\\
&&{\cal C}=\frac{2}{d+1}[d(\rho +\tau ) +(d^2+d-4)t]\ . 
\la{C}
\eea
A straightforward calculation of the correlators gives 
$$\tau =-\l\frac{k+1}{\omega_{d-1}}\r^3\cdot\frac{(2k)!}{(k!)^2}, \quad \quad 
t =- 2 \l\frac{k+1}{\omega_{d-1}}\r^3\cdot\frac{(2k-2)!}{k!(k-1)!},
\quad \quad \rho =0\  .$$
Using formulas (\ref{A})--(\ref{C}) we thus
obtain the values of 
the three independent constants
\bea
&&{\cal A}= -\frac{8}{2k-1}
\l\frac{k+1}{\omega_{2k+1}}\r^3\cdot n_{ph}
\la{Ah}\\
&&{\cal B}=-\frac{16 k}{2k-1}
\l\frac{k+1}{\omega_{2k+1}}\r^3\cdot  n_{ph}
\la{Bh}\\
&&{\cal C}=-\frac{8}{2k-1}
\l\frac{k+1}{\omega_{2k+1}}\r^3\cdot  n_{ph}
\la{Ch}
\eea
A check of these  formulas is provided by the conformal Ward identity 
relating the 2-point and 3-point correlators \cite{EO}
\bea
C_T=\frac{\omega_{d-1}}{d(d+2)}\left[ \frac12 (d+2)(d-1){\cal A}-{\cal B}-
2(d+1){\cal C}\right] \ . 
\la{ward}
\eea
In terms of $t,~\tau ,~\rho$ this  identity can be rewritten in the form
\bea
C_T=-4\omega_{d-1}\left( \frac{1}{d+2}\rho + \frac{1}{d}\tau \right) \ . 
\nonumber
\eea
One can easily check the validity of this  equation using 
(\ref{ct}). It is worth noting 
that the conformal Ward identity relation  does not involve 
the constant  $t$.

In the case of an even-rank $k=2l$ field  $B$  the stress 
tensor (\ref{T}) admits factorization into the sum of
the  ``left" 
and ``right" stress tensors depending on the anti selfdual and selfdual
components of the field $H=dB$ respectively:
\bea
T_{\a\b}=T_{\a\b}^{-}+T_{\a\b}^{+}\ ,\ \ \ \ \ \ \ \ \ \ \ \
T_{\a\b}^{\pm}=\frac{1}{k!}  H^\pm_{\a\mu\nu}H_\b^{\pm\mu\nu}\  . 
\la{fact}
\eea
Moreover, one can easily show that the correlator of $H^-$ and $H^+$ 
is proportional to the delta-function. 
We assume a regularization were contact terms proportional to 
delta functions are omitted, as this is consistent with 
conformal invariance, 
and so we can set both  $\langle H^-(x) H^+(y) \rangle$ 
and $\langle T_{\a\b}^{-}(x) T_{\g\d}^{+}(y) \rangle$ to zero.
This  serves as a justification for  the  following 
prescription (essentially equivalent to  the one originally 
used in  \cite{AW} and also in \cite{GKT})   to 
compute the  correlation functions of  operators depending 
on the field strength 
 in a chiral model:
 one may  use  the non-chiral
propagator (\ref{prop})  while  replacing    the field 
strength $H$ by 
its (anti)selfdual part in the  composite operators. 
In the case of the 2-point  and 3-point correlation functions of 
the stress tensor 
$T_{\a\b}^{\pm}$  the correlator of  
$T_{\a\b}^{-}$ and $T_{\a\b}^{+}$ vanishes, and thus 
the  chiral correlators  are equal to  $1 \over 2$  of 
the 2-point  and 3-point ``non-chiral" correlators  of the full
 $T_{\a\b}$.

\section{The (2,0) tensor multiplet in $d=6$}
The (2,0) tensor multiplet in the 6-dimensional Minkowski space 
consists of 5 scalars $X^i$, 2 Weyl fermions $\psi^I_L$  and an 
antisymmetric tensor $B_{\a\b}$ with anti-selfdual strength. 
Covariant lagrangian descriptions of the 2-form part of the model  
exist \cite{PST}, but are hard  to work with 
at the quantum level,  
since one cannot easily implement a covariant gauge 
fixing for the gauge 
symmetry that gives the (anti)selfduality  constraint.
However, it is sufficient for our aims to use a Lagrangian
containing a non-chiral 2-form with the prescription of projecting
out its selfdual part in the relevant composite operators.
Thus, the stress tensor of the system is given by the sum of the stress tensors
of the fields
\be
T_{\a\b}=T_{\a\b}^{H^-} +T_{\a\b}^X+T_{\a\b}^\psi\  ,
\label{coi}
\ee
with
\bea
\la{th}
T_{\a\b}^{H^-}&=&\frac12 H^-_{\a\mu\nu}H_\b^{-\mu\nu}
\\
\la{tx}
T_{\a\b}^X&=&\p_\a X^i\p_\b X^i-\frac{1}{5}\p_\a\p_\b (X^iX^i)
-\frac{1}{10}\eta_{\a\b}\p_{\mu}X^i\p^\mu X^i \\
\la{tpsi}
T_{\a\b}^\psi&=&-\frac i4 \bar{\psi}^I_L(\g_\a\p_\b +\g_\b\p_\a )\psi^I_L
+\frac i4 (\p_\b\bar{\psi}^I_L\g_\a\psi^I_L+\p_\a\bar
{\psi}^I_L\g_\b\psi^I_L)  \ , 
\eea
where $i=1,...,5$ and $\psi^I_L$   is the
left component of a Dirac fermion.
We took into account that the stress tensor for $H$ can be 
represented in the form
$$T_{\a\b}^H=  \frac12 H_{\a\mu\nu} H_\b^{\ \mu\nu}
-\frac{1}{12}\eta_{\a\b} H_{\mu\nu\rho}^2 = 
T_{\a\b}^{H^-}+T_{\a\b}^{H^+}\ .$$
The stress tensor  
(\ref{coi})
 coincides,  up to a  factor,   with 
the stress tensor for the (2,0) tensor multiplet found  
in \cite{BSV} by using a different method.
Note that the scalar stress tensor contains 
the improvement term as needed for conformal invariance
(on-shell tracelessness).\footnote{The improvement term 
originates from the ${(d-2)\over 4(d-1)}R X^2$ term on a 
curved $d=6$ background.
This term was not included in the absorption calculation in \cite{GKT}
since it gives zero contribution to the tree-level 3-point 
amplitude  with on-shell graviton. However, this term 
is crucial for the  scalar stress tensor correlators to have the 
canonical CFT form described in the previous section.}
 
The  2- and 3-point correlation functions of the stress tensors of the 
free scalar and  spinor  theories in arbitrary number $d$ of dimensions 
 were previously   computed 
in \cite{OP}.  We extend those results by including the
 contributions
of  $k$-forms (with the understanding 
that this additional contribution
is present  only in the suitable ``conformal'' 
$d=2k+2$ dimensions). 
Thus,
the corresponding
constants $C_T$,  ${\cal A}$, ${\cal B}$ and ${\cal C}$  in $<TT>$
(\ref{two}) and $<TTT>$
(\ref{three}) are given by
\bea
 C_T&=&\frac{1}{\omega_{d-1}^2}\l \frac{d}{d-1}\ n_S +\ \frac{d}{2} \ 
\tilde{n}_F +\ \frac{d^2}{2}\ \tilde{n}_{B}\r \nonumber\\
{\cal A}&=& -\frac{1}{\omega_{d-1}^3} \left[ -\frac{d^3}{(d-1)^3}n_S
+\ \frac{d^3}{d-3} \tilde{n}_{B} \right]
\nonumber\\
{\cal B}&=&-\frac{1}{\omega_{d-1}^3}\left[ \frac{(d-2)d^3}{(d-1)^3}n_S
+\frac{d^2}{2} \tilde{n}_F+\frac{(d-2)d^3}{d-3} \tilde{n}_{B}
\right]\nonumber\\
{\cal C}&=&-\frac{1}{\omega_{d-1}^3}\left[ \frac{(d-2)^2d^2}{4(d-1)^3}n_S
+\frac{d^2}{4}\tilde{n}_F + \frac{d^3}{d-3} \tilde{n}_{B}
\right] \ . 
\la{ctbf}
\eea
Here $\tilde{n}_F=  \tr I\cdot  n_F $ (tr is the Dirac spinor trace)  and
$\tilde n_B = \frac{(2k)!}{(k!)^2}\ n_B$, with 
$n_S$, $n_F$ and $n_B$ the numbers of scalars, Dirac spinors 
and $k$-forms, respectively. For Weyl fermions and 
chiral $k$-forms one should halve the corresponding 
numbers\footnote{Except in $d=2$
where the $\epsilon$ tensor gives a new structure in the
stress tensor 2-point function and the chiral splitting happens in
a different way.}.

Summing up the contributions of the anti-selfdual tensor, 5 scalars and 2 
Weyl fermions, we  finally 
obtain the  following values of the four basic constants
$C_T$,  ${\cal A}$, ${\cal B}$ and ${\cal C}$ for the (2,0) 
tensor multiplet in $d=6$ 
\be
C_T=\frac{84}{\pi^6}\ ,\quad\ \la{free} 
\ee
\bea
{\cal A}=-\frac{2^6\cdot 3^4}{5^2\pi^9}\ ,\quad\ \ \   
{\cal B}=-\frac{181\cdot 2^4\cdot 3^2}{5^2\pi^9}\ ,\quad\ \ \ 
{\cal C}=-\frac{59\cdot 2^3\cdot 3^3}{5^2\pi^9}\ . 
\la{Cf}
\eea
We are now going to compare these 
{\it  free field theory}  results with 
the ones obtained  from 
 the 11-dimensional supergravity 
on the $AdS_7\times S^4$ background 
describing the near-horizon limit of 
$N$ coincident M5-branes. In units in which 
the radii of the two spaces are 
$R_{AdS}=1$ and
$R_{S^4}={1\over 2}$
 the 11-dimensional gravitational constant 
 ($S= - \frac{1}{2\kappa^2_{11}} \int d^{11} x \sqrt g R + ...$) 
 is \cite{kt,kttt,gktt} 
\be
\frac{1}{2\kappa^2_{11}}=\frac{2N^3}{\pi^5}\ .
\ee
Performing the dimensional reduction to seven dimensions, we get the 
7-dimensional gravitational constant (Vol$(S^4) = \omega_4 ({1 \over 2})^4$)
\be 
\frac{1}{2\k_7^2}=\frac{N^3}{3\pi^3}\  .
\ee 
In general, the  constant $C_T$ in (\ref{two})  calculated by using the  
$AdS_{d+1}$ supergravity description  is given by \cite{LT}
\bea
C_{T}^{(ads)}=\frac{1}{2\k_{d+1}^2}\cdot \frac{2d(d+1)
\Gamma (d)}{(d-1)\Gamma ({d/2})\pi^{d/2}} \ . 
\la{cads}
\eea
We assumed that the coupling of $T_{\a\b}(x)$ with $h_{\a\b}(x)$   at  the 
boundary of $AdS_{d+1}$  has the standard form  
$\int d^dx\ \frac{1}{2}T_{\a\b}(x)h^{\a\b}(x)  $.
Taking into account the value of the gravitational constant $\k_7$, we find
that for $d=6$ 
\bea
C_T^{(ads)}=4N^3\cdot\frac{84}{\pi^6}.
\la{cads6}
\eea
This differs by the factor $4N^3$ from the value  (\ref{free}) 
obtained from  
the free field theory. This factor coincides with the one 
obtained in \cite{GKT} by 
comparing the 
absorption cross-sections calculated using the $d=11$ supergravity and 
the free 
world-volume field theory descriptions.
 The constants 
${\cal A}$, ${\cal B}$ and ${\cal C}$ were computed in the general case of 
$AdS_{d+1}$ gravity in \cite{AF}. For $d=6$ they are given by 
\bea
{\cal A}^{(ads)} &=& -4N^3\cdot {2^6 \cdot 3^4 \over 5^2 \pi^9}\ ,
\cr 
{\cal B}^{(ads)} &=& -4N^3\cdot {181\cdot 2^4
 \cdot 3^2 \over 5^2 \pi^9}\ ,\cr 
{\cal C}^{(ads)} &=& -4N^3\cdot 
{59\cdot 2^3 \cdot 3^3 \over 5^2 \pi^9\ }\ .
\eea
Comparing these values with the ones obtained in the (2,0) tensor 
multiplet model (\ref{Cf}) we see that they again 
differ by the same factor $4N^3$.  
The overall scale of these three constants 
is, of course, determined by $C_T$ in view of the 
conformal Ward identity (\ref{ward}),
but it is quite 
remarkable that the {\it relative} scales
of $\cal A, B, C$ are exactly the same in the 
free field theory and in the $AdS_7$ supergravity!

This suggests  that all  2- and 3-point correlation functions
of the states from the stress tensor short multiplet 
in the  interacting (2,0) superconformal  field theory 
coincide (in the  large $N$ limit)
  with the  corresponding ones in  the free theory of $4N^3$ (2,0) 
tensor multiplets.
 As a  check  of this expectation   we  can calculate
 the 
2- and 3-point functions of the lowest chiral primary operators
\be
{\cal O}^I = {\cal C}^I_{ij} X^iX^j\ ,  
\la{cpo}
\ee
where ${\cal C}^I_{ij}$ is a symmetric traceless tensor, 
while the index $I$ denotes a complete basis of such tensors, and 
compare them with the ones obtained from the 11-dimensional
supergravity in \cite{BZ}.  Normalizing the operators  in such a way that the 
ratio of the 2-point functions calculated in $AdS_7$  gravity and in 
the free theory is $4N^3$, we find  again 
that the ratio of the 3-point
functions is also $4N^3$.

\section{M2-brane  case: $AdS_4$ -- free  $d=3$ CFT comparison }
In this section we compare the 2- and 3-point 
functions of the stress tensor of the effective action for 
$N$ M2-branes calculated from the 11-dimensional 
$AdS_4\times S^7$ supergravity and in the 3-dimensional 
free field theory  of
8 scalars and 8 Majorana fermions. 

We choose units in which $R_{AdS}=1$, so  that  
 $R_{S^7}=2$, and  thus  
\be {1\over 2\kappa^2_{11}} = {N^{3\over2} \over 2^9 \sqrt{2} 
 \pi^5},\quad\quad\quad\ \ \ \ \ 
{1\over 2\k_{4}^2} = {N^{3\over2} \over 12 \sqrt{2} \pi}\ .\ee
Then from (\ref{cads}) and \cite{AF} we get
\be
C_T^{(ads)} ={4\over \pi^3}\cdot {N^{3\over 2}\over \sqrt 2}\ ,
\la{ccc}
\ee
\bea
{\cal A}^{(ads)} = - {27\over 8 \pi^4} 
\cdot{N^{3\over 2}\over \sqrt 2}\ ,\quad\quad
{\cal B}^{(ads)} = - {57\over 8 \pi^4} 
\cdot {N^{3\over 2}\over \sqrt 2}\ ,\quad\quad
{\cal C}^{(ads)} = - {99\over 32 \pi^4} 
\cdot {N^{3\over 2}\over \sqrt 2}\ . 
\la{cabcads}
\eea
By using (\ref{ctbf}) we obtain the following values of the 4 basic
constants in the free $d=3$ conformal  field 
theory of 8 scalars and 8 Majorana fermions 
\be
C_T^{(free)} ={3\over 2\pi^2}\ ,\la{cccc}
\ee
\bea 
{\cal A}^{(free)} = {27\over 64 \pi^3}\ ,\quad\quad
{\cal B}^{(free)} = - {63\over 64 \pi^3}\ ,\quad\quad
{\cal C}^{(free)} = - {81\over 256 \pi^3}\ . 
\la{cabcfree}
\eea
Although the constants (\ref{cabcads}) and (\ref{cabcfree}) look different, 
this  does not mean that the corresponding 3-point functions differ too. 
In fact, as was  shown in \cite{OP},  in  {\it three}
dimensions there are only {\it two} 
independent conformal tensor structures  in $<TTT>$, and, therefore, 
only two linear combinations of the constants in (\ref{three}) 
have got an invariant meaning.
These two independent  constants may 
be expressed  in terms of 
${\cal A}$, ${\cal B}$ and ${\cal C}$ as follows
\be 
{\cal P} =  4 {\cal A} +3 {\cal B} -14 {\cal C}\ ,\quad\quad 
{\cal Q} = {\cal A} -2 {\cal C}\ . \ee
Then a straightforward calculation gives
\bea
{C_T^{(ads)}\over C_T^{(free)}}  =  
{{\cal P}^{(ads)} \over {\cal P}^{(free)}} = 
{{\cal Q}^{(ads)} \over {\cal Q}^{(free)}} = 
{4 \sqrt{2} \over 3 \pi} N^{3\over 2}  \ .  
\la{ratio}
\eea

Using the results of \cite{BZ} and a simple free theory computation,
one can also check that the ratio of the 2- and 3-point functions
of the properly normalized chiral primary operators (\ref{cpo}) 
is again given by the same factor (\ref{ratio}), which coincides 
also with the one obtained in the comparison of graviton 
absorption cross-sections
in \cite{GKT}.
 It  
seems natural to expect  that 
all 2- and 3-point functions of operators from the short multiplet
of the stress tensor 
in the effective theory of $N$\  M2-branes
coincide, up to this overall factor, 
 with the ones computed  in the ${\cal N}=8$ free field theory.
 The meaning of this irrational proportionality constant
   (which   looks somewhat ugly 
compared to $4N^3$ in the $d=6$ case)
remains unclear.

\vskip 5cm
{\bf Acknowledgements}

 F.B. would like to thank R. Zucchini, and 
A.T. would like to thank A. Hashimoto, C. Hull,  
I. Klebanov and H. Liu for useful  discussions. 
The work of S.F. was supported by
the U.S. Department of Energy under grant No. DE-FG02-96ER40967.
The work of A.T. was  supported in part by
the U.S. Department of Energy under grant No. DOE/ER/01545-775, 
by the EC TMR programme ERBFMRX-CT96-0045, 
INTAS grant No.96-538,
and NATO grant PST.CLG 974965.

\baselineskip=14pt

\end{document}